\documentclass[a4paper,11pt]{article}
\pdfoutput=1
\usepackage{jcappub}
\usepackage{graphicx}
\usepackage{amsmath,amssymb}
\usepackage{float}
\usepackage{lineno}
\nolinenumbers
\graphicspath{{./}}
\makeatletter\gdef\@fpheader{}\makeatother
\begin{document}
\pagestyle{empty}
\title{\boldmath Beyond the equation of state: a second-order diagnostic for dynamical dark energy}

\author{B.~Osano}
\affiliation{Cosmology and Gravity Group, Department of Mathematics
and Applied Mathematics, and Centre for Higher Education Development,\\
University of Cape Town, Rondebosch 7701, South Africa}
\emailAdd{bob.osano@uct.ac.za}

\abstract{
The first-order continuity equations determine the evolution of the energy densities but depend only on the instantaneous value of the dark-energy equation-of-state parameter. Differentiating these equations with respect to e-fold time introduces the term $\omega'_{\rm DE}$ explicitly, providing a second-order probe of dark-energy dynamics. Consequently, while information about the evolution of the equation of state is encoded in the full dynamical solution, it is not explicit in the first-order continuity equations evaluated at a given epoch. The second-order formulation, therefore, provides a complementary description in which the local evolution of the equation of state appears directly through the curvature of the density trajectory.

For a two-fluid interacting dark-sector model with linear coupling $Q_{AB}=\alpha\rho_AH$, the resulting second-order equation defines a curvature diagnostic, $\mathcal{C}=\rho_{DE}''/\rho_{DE}$, whose leading contribution, in the cosmological-constant limit, is $\alpha^2$, while departures from $\omega_{DE}=-1$ generate corrections through both $\delta\omega=1+\omega_{DE}$ and the distinctive term $-3\omega_{DE}'$. Unlike first-order analyses, this contribution is independent of the interaction strength and directly identifies dynamical dark energy. Applying the diagnostic to a CPL model with parameters consistent with DESI constraints, we recover $\omega_{DE}'$ across the full redshift range for both weak and strong interactions. Noise propagation shows that the diagnostic is detectable with signal-to-noise ratio exceeding three for $\sigma_H/H\lesssim1.5\%$, while the degeneracy between $\alpha$ and $\omega_{DE}'$ remains negligible for $\alpha\lesssim0.1$. In the non-interacting limit, the formalism naturally recovers the Caldwell--Linder thawing/freezing classification and extends it to interacting dark-energy models.
}

\maketitle

\section{\label{sec:intro}Introduction}

A central question in modern cosmology is whether the dark-energy equation of state is truly constant or evolves with cosmic time. Although the concordance $\Lambda$CDM model successfully describes a wide range of cosmological observations by assuming a cosmological constant with $w=-1$ \cite{Peebles2003,Weinberg1989}, this assumption is phenomenological rather than derived from first principles. A broad class of dynamical dark-energy models therefore remains viable \cite{Copeland2006,Frieman2008,Weinberg2013}. Interest in such models has been strengthened by recent analyses of the first and second data releases of the Dark Energy Spectroscopic Instrument (DESI), which, when combined with cosmic microwave background and Type Ia supernova observations, show a mild preference for an evolving dark-energy equation of state over the baseline $\Lambda$CDM model \cite{DESI}.

A natural extension of $\Lambda$CDM is provided by interacting dark-sector models, in which dark matter and dark energy exchange energy while the total energy-momentum tensor remains conserved \cite{Amendola2000,Zimdahl2001,Wang2016}. Such models have been widely investigated as possible resolutions of the coincidence problem and the Hubble tension, and more recently as viable descriptions of the DESI data \cite{Bolotin2015,Wang2016,Giare2024}. Their defining feature is the interaction term appearing in the continuity equations, which modifies the background evolution and introduces a degeneracy between energy transfer and intrinsic dark-energy dynamics. Consequently, the equation of state inferred from cosmological observations is generally an effective quantity that need not coincide with the underlying microscopic equation of state \cite{Valiviita2008,Kunz2009}.

Most existing diagnostics of interacting dark energy are based on the first-order continuity equations or on integrated background observables. While these approaches constrain the effective evolution of the dark sector, they do not distinguish between the instantaneous value of the equation-of-state parameter and its rate of evolution. Within the dynamical-systems framework, however, the derivative $\omega_{DE}'$ plays a fundamental role. Caldwell and Linder showed that the $w$--$w'$ phase plane separates quintessence models into thawing and freezing classes, establishing $\omega_{DE}'$ as a key indicator of dark-energy dynamics \cite{Wainwright1997,Coley2003,Caldwell2005,Linder2006,Scherrer2008}. Nevertheless, this diagnostic was developed for non-interacting scalar-field models and has not been derived directly from the continuity equations.

In this paper, we show that differentiating the continuity equations with respect to e-fold time naturally generates a term proportional to $\omega_{DE}'$, which is absent from the first-order formulation. The resulting second-order continuity equations provide a higher-order dynamical diagnostic that is directly sensitive to the evolution of the dark-energy equation of state while remaining independent of any particular scalar-field potential. That differentiating a continuity equation containing $\omega$ as a coefficient will produce $\omega'$ is, in itself, a straightforward consequence of the product rule. What is non-trivial is the structural outcome: the term $-3\omega_{DE}'$ emerges as an isolated, interaction-independent contribution to the curvature coefficient, distinct from all terms involving $\alpha$, and with no counterpart at first order regardless of the interaction strength. To the best of our knowledge, this structural isolation and its extension to interacting dark-sector cosmologies have not been identified previously. Our formulation therefore provides a dynamical derivation of the Caldwell--Linder $w$--$w'$ diagnostic and extends it to models with non-vanishing dark-sector interactions. It also complements our earlier dynamical-systems studies of the matter--dark-energy transition in interacting cosmologies \cite{OsanoOreta2019,Osano2025,Osano2025b}.

The remainder of this paper is organised as follows. Section~\ref{sec:setup} introduces the interacting two-fluid framework, while Section~\ref{sec:derivation} derives the second-order continuity equations. In Section~\ref{sec:de} we specialise the formalism to the dark matter--dark energy system, examine the cosmological-constant limit, analyse the hierarchy of interaction corrections, illustrate the method using the Chevallier--Polarski--Linder parametrisation, and assess the observational viability of the proposed diagnostic. Section~\ref{sec:discussion} discusses the implications of the results in the context of the Caldwell--Linder classification and recent DESI observations, and Section~\ref{sec:conclusions} presents our conclusions.
\section{\label{sec:setup}Two-fluid interacting dark-sector model}

\subsection{Spacetime, fluids, and interaction}

We consider a spatially flat Friedmann--Lema\^{\i}tre--Robertson--Walker (FLRW) spacetime with line element
\[
ds^2=-dt^2+a^2(t)\delta_{ij}dx^idx^j,
\]
where $a(t)$ denotes the scale factor and $H\equiv\dot a/a$ is the Hubble expansion rate; an overdot represents differentiation with respect to cosmic time $t$ \cite{Weinberg1972,Mukhanov2005}. Throughout this paper, we adopt the e-fold number,
\[
N\equiv\ln a,
\]
as the independent time variable. The e-fold parametrisation provides a natural measure of cosmological evolution and recasts the background equations into a form that is particularly well suited to dynamical analyses \cite{Copeland1998,Bahamonde2018}.

We assume that the cosmic fluid consists of two interacting barotropic components, labelled $A$ and $B$, together with a separately conserved component $C$ representing radiation. Each fluid satisfies the barotropic equation of state
\[
p_i=\omega_i(N)\rho_i,
\]
where $\rho_i$ and $p_i$ denote the energy density and pressure of the $i$th fluid, respectively. We allow the equation-of-state parameter $\omega_i$ to be an arbitrary, sufficiently smooth function of the e-fold number and make no assumption regarding its functional form. Consequently, the analysis remains applicable to a broad class of interacting dark-energy models.

The barotropic assumption enables us to eliminate the pressure from the continuity equations by expressing it algebraically in terms of the energy density. As a result, we obtain a closed system of evolution equations involving only the energy densities and their derivatives with respect to $N$. This formulation provides the foundation for the second-order continuity equations derived in the following section.

We assume that the total energy-momentum tensor is covariantly conserved,
\begin{equation}
\nabla_{\mu}\!\left(\sum_{i=A,B,C}T^{\mu\nu}_{i}\right)=0.
\label{eq:totalcons}
\end{equation}  We allow, however, for an exchange of energy between the dark-sector components. Following the standard phenomenological treatment of interacting dark-energy models \cite{Amendola2000,Zimdahl2001,Wang2016}, we describe the interaction by an energy-transfer term $Q_{AB}$, where the subscripts denote the interacting fluids. Throughout this work, we adopt the convention that $Q_{AB}>0$ corresponds to a net transfer of energy from fluid $A$ to fluid $B$. The third component, $C$, representing radiation, is assumed to be separately conserved and therefore does not participate in the interaction.

We adopt the linear interaction ansatz
\begin{equation}
Q_{AB} = \alpha\,\rho_A\,H,
\label{eq:Qansatz}
\end{equation}
where $\alpha$ is a dimensionless coupling constant. This choice
preserves the correct dimensions, ensures that the interaction
vanishes when the source fluid $A$ is absent, and is the most
widely studied phenomenological form in the interacting dark-energy
literature \cite{Amendola2000,Wang2016,Yang2018,Avelino2009}. We take $\alpha$ to be
constant throughout; the consequence of relaxing this assumption for
the second-order equations is discussed below.

Converting to e-fold time using $\dot{f}=Hf'$, where a prime denotes
$d/dN$, and dividing equation~(\ref{eq:totalcons})
through by $H$, the ansatz~(\ref{eq:Qansatz}) yields
\begin{eqnarray}
\rho_A' &=& -\bigl[3(1+\omega_A)+\alpha\bigr]\,\rho_A ,
\label{eq:rhoAfirst}\\[4pt]
\rho_B' &=& -3(1+\omega_B)\,\rho_B + \alpha\,\rho_A ,
\label{eq:rhoBfirst}\\[4pt]
\rho_C' &=& -3(1+\omega_C)\,\rho_C .
\label{eq:rhoCfirst}
\end{eqnarray}
Equation~(\ref{eq:rhoAfirst}) shows that the interaction augments
the standard Hubble-dilution rate of fluid $A$ by the constant $\alpha$;
equation~(\ref{eq:rhoBfirst}) shows that fluid $B$ gains a source
proportional to $\rho_A$. The three equations decouple hierarchically:
$\rho_A$ can be solved independently, after which $\rho_B$ is
determined by a driven equation with $\rho_A$ as source, and $\rho_C$
evolves independently throughout.

Three assumptions govern the derivations in Section~\ref{sec:derivation}.
(i)~$\alpha'=0$: the coupling is constant. If $\alpha$ were a function
of $N$, differentiating equation~(\ref{eq:rhoAfirst}) would generate
an additional term $-\alpha'\rho_A$ in the second-order equation for
fluid $A$ alongside the $\omega_A'$ term; the analogous correction
$-\alpha'\rho_A$ would also appear in the source term of the
second-order equation for fluid $B$. The appearance of $\omega_i'$
at second order is qualitatively unchanged, but the full expression
becomes more complex.
(ii)~$\omega_A$ and $\omega_B$ are smooth functions of $N$ with no
further restriction.
(iii)~Spatial flatness holds, so equation~(\ref{eq:totalcons}) reduces
to the standard FLRW constraint.

\subsection{Expansion-normalised variables and background equations}

For the dynamical-systems analysis that the second-order equations
will motivate, it is convenient to work with the dimensionless
expansion-normalised density parameters \cite{Bahamonde2018}
\begin{equation}
X \equiv \Omega_A = \frac{\rho_A}{3H^2}, \qquad
Y \equiv \Omega_B = \frac{\rho_B}{3H^2}, \qquad
Z \equiv \Omega_C = \frac{\rho_C}{3H^2},
\label{eq:densparams}
\end{equation}
so that the first Friedmann equation reduces, under assumption~(iii)
of Section~\ref{sec:setup} (spatial flatness), to the algebraic constraint
\begin{equation}
X + Y + Z = 1.
\label{eq:friedmann}
\end{equation}
The effective equation of state of the combined system, which controls
the deceleration parameter $q = -1 - H'/H$, is
\begin{equation}
\omega_{\rm eff} = \sum_i \omega_i\,\Omega_i
= \omega_A X + \omega_B Y + \omega_C Z.
\label{eq:omegaeff}
\end{equation}
This weighted average over the individual equations of state, not any
single $\omega_i$, is what appears in the background expansion history
inferred from distance measurements. In the interacting dark-sector
setting, where energy transfer modifies the individual $\rho_i$
trajectories, $\omega_{\rm eff}$ can differ substantially from
$\omega_{DE}$ even when $\omega_{DE}$ is static \cite{Avelino2009}.

The complete autonomous background system also requires the
Raychaudhuri equation, which in e-fold time reads
\begin{equation}
\frac{H'}{H} = -\frac{3}{2}\bigl[(1+\omega_A)\Omega_A
+ (1+\omega_B)\Omega_B + \tfrac{4}{3}\Omega_C\bigr],
\label{eq:Hprime}
\end{equation}
where the factor $4/3$ in the radiation term reflects $\omega_C=1/3$.
Together, equations~(\ref{eq:rhoAfirst})--(\ref{eq:rhoCfirst}),
(\ref{eq:friedmann}), and~(\ref{eq:Hprime}) form the complete
background system. For the second-order derivations of
Section~\ref{sec:derivation}, only
equations~(\ref{eq:rhoAfirst})--(\ref{eq:rhoBfirst}) and the
constancy of $\alpha$ are needed directly. Equation~(\ref{eq:Hprime})
enters when the system is rewritten in $\Omega_i$ variables, as
discussed in Section~\ref{sec:derivation}, and would also appear if
the coupling were time-varying.

\section{\label{sec:derivation}Second-order density evolution}

The first-order equations~(\ref{eq:rhoAfirst})--(\ref{eq:rhoBfirst})
determine the slope of each density trajectory in the $(N,\rho)$-plane
at every instant. They carry information about $\omega_i$ as a
coefficient, but not about whether $\omega_i$ is changing. To access
that information, we differentiate once more with respect to $N$.
The calculation is straightforward but the result is structurally
significant: the second derivative of each density contains a term
proportional to $\omega_i'$ that has no counterpart at first order,
and the two equations become coupled in a new way through the
cross-term in the equation for fluid $B$.

\subsection{Second-order equation for fluid $A$}

Apply $d/dN$ to equation~(\ref{eq:rhoAfirst}). Since $\alpha$ is
constant by assumption~(i) of Section~\ref{sec:setup}, and $\omega_A$
is a smooth function of $N$ by assumption~(ii), the product
rule on the right-hand side yields two contributions: differentiating
the coefficient $[3(1+\omega_A)+\alpha]$ generates $3\omega_A'$,
and differentiating $\rho_A$ reintroduces $\rho_A'$:
\begin{equation}
\rho_A'' = -3\omega_A'\,\rho_A
           -\bigl[3(1+\omega_A)+\alpha\bigr]\,\rho_A'.
\label{eq:rhoA_diff}
\end{equation}
Substituting equation~(\ref{eq:rhoAfirst}) for $\rho_A'$ and
collecting terms:
\begin{equation}
\rho_A'' = \left\{\bigl[3(1+\omega_A)+\alpha\bigr]^2
           - 3\omega_A'\right\}\rho_A.
\label{eq:rhoAsecond_compact}
\end{equation}
This compact form has a transparent structure: the squared bracket
is the square of the first-order coefficient from
equation~(\ref{eq:rhoAfirst}), and the term $-3\omega_A'$ is the
correction that arises solely from differentiating that coefficient.
When $\omega_A$ is constant, $\omega_A'=0$ and the second-order
equation contains only the square of the first-order coefficient.
When $\omega_A$ evolves, $\omega_A'\neq 0$ and a genuinely new term
appears that has no counterpart at first order. Expanding the square
gives the explicit form
\begin{equation}
\rho_A'' = \left[9(1+\omega_A)^2 + 6\alpha(1+\omega_A) + \alpha^2
           - 3\omega_A'\right]\rho_A .
\label{eq:rhoAsecond}
\end{equation}

The four terms in equation~(\ref{eq:rhoAsecond}) each have a distinct
physical origin. The leading term $9(1+\omega_A)^2$ is the square of
the non-interacting dilution rate $3(1+\omega_A)$ and is present even
when $\alpha=0$; it represents the curvature induced by background
expansion alone. The terms $6\alpha(1+\omega_A)$ and $\alpha^2$ are
linear and quadratic in the coupling respectively; they measure the
effect of energy transfer on the curvature of the trajectory, with the
quadratic term surviving even in the cosmological-constant limit
$\omega_A=-1$. The term $-3\omega_A'$ is the diagnostic term: it
encodes the rate of change of the equation of state and is entirely
absent from the first-order equation~(\ref{eq:rhoAfirst}), where
$\omega_A$ enters only as a coefficient, not its derivative.

\subsection{Second-order equation for fluid $B$}

Apply $d/dN$ to equation~(\ref{eq:rhoBfirst}). The right-hand side
has two terms: the self-dilution term $-3(1+\omega_B)\rho_B$ and the
interaction source $+\alpha\rho_A$. Differentiating each:
\begin{equation}
\rho_B'' = -3\omega_B'\rho_B - 3(1+\omega_B)\rho_B' + \alpha\rho_A'.
\label{eq:rhoB_diff}
\end{equation}
Substituting equation~(\ref{eq:rhoBfirst}) for $\rho_B'$ and
equation~(\ref{eq:rhoAfirst}) for $\rho_A'$, and separating
$\rho_B$ and $\rho_A$ terms:
\begin{eqnarray}
\rho_B'' &=& \bigl[9(1+\omega_B)^2 - 3\omega_B'\bigr]\rho_B  -\; \alpha\bigl[3(1+\omega_B) + 3(1+\omega_A) + \alpha\bigr]\rho_A.
\label{eq:rhoB_intermediate}
\end{eqnarray}
Combining $3(1+\omega_B)+3(1+\omega_A) = 3(2+\omega_A+\omega_B)$
gives the final form
\begin{equation}
\rho_B'' = \left[9(1+\omega_B)^2 - 3\omega_B'\right]\rho_B
           - \alpha\left[3(2+\omega_A+\omega_B)+\alpha\right]\rho_A .
\label{eq:rhoBsecond}
\end{equation}

Two features of equation~(\ref{eq:rhoBsecond}) merit attention.
First, like equation~(\ref{eq:rhoAsecond}) for fluid $A$, it contains
the diagnostic term $-3\omega_B'$; the curvature of fluid $B$'s
trajectory is also sensitive to whether its own equation of state
is evolving. Second, the cross-term $-\alpha[3(2+\omega_A+\omega_B)+\alpha]\rho_A$
is absent from equation~(\ref{eq:rhoBfirst}) at first order.
The coefficient $3(2+\omega_A+\omega_B) = 3(1+\omega_A)+3(1+\omega_B)$
is the sum of the effective pressure contributions of both fluids; the cross-term therefore couples the curvature of fluid $B$'s
trajectory to the dynamical state of both $A$ and $B$ simultaneously.
This coupling is entirely a consequence of the interaction and vanishes
identically when $\alpha\to 0$.

\subsection{\label{sec:structure}Structural properties of the second-order system}

Equations~(\ref{eq:rhoAsecond}) and~(\ref{eq:rhoBsecond}) together
constitute the second-order description of the interacting system.
Two structural properties distinguish the second-order system from
the first-order system and are the basis of the diagnostic argument.

\textit{Property 1: sensitivity to $\omega_i'$.}
Both equations depend explicitly on $\omega_A'$ and $\omega_B'$.
In the first-order system~(\ref{eq:rhoAfirst})--(\ref{eq:rhoBfirst}),
each equation of state appears only as a coefficient of $\rho_i$.
Evaluating the right-hand side of equation~(\ref{eq:rhoAfirst}) at
two models that share the same instantaneous $\omega_A$ but differ in
$\omega_A'$ produces identical results; they are locally
indistinguishable at first order. The same evaluation of
equation~(\ref{eq:rhoAsecond}) produces different results because the
$-3\omega_A'$ term is different. The second-order equations therefore
carry information about the evolution of the dark-energy equation of
state that the first-order equations structurally cannot.

\textit{Property 2: coupling structure.}
In $\rho_i$ variables the second-order equations mirror the
hierarchical decoupling of the first-order system: $\rho_A''$
depends only on $\rho_A$ and $\omega_A$, while $\rho_B''$ depends
on $\rho_A$, $\rho_B$, $\omega_A$, and $\omega_B$. When the system
is rewritten in the expansion-normalised variables
$\Omega_i = \rho_i/3H^2$, however, the identity
$\Omega_i' = \rho_i'/(3H^2) - 2(H'/H)\Omega_i$ brings in the
Raychaudhuri equation~(\ref{eq:Hprime}), which couples $H'/H$ to the
total gravitating density. In $\Omega_i$ variables all three components
therefore couple to each other, and the decoupling of the $\rho_i$
system is broken. This global coupling in $\Omega_i$ variables is
the relevant one for the critical-point analysis of
Section~\ref{sec:discussion}, where the expansion-normalised
variables are the natural phase-space coordinates.

Together, Properties 1 and 2 mean that the second-order system
provides qualitatively new information at two levels: within each
fluid's own evolution through $\omega_i'$, and across the system
as a whole through the cross-terms and the $\Omega_i$ coupling. The
application to dark energy in Section~\ref{sec:de} will show how
Property~1 yields a concrete diagnostic for whether dark energy
is dynamical.

\section{\label{sec:de}Application to interacting dark energy}

The general second-order equations~(\ref{eq:rhoAsecond})
and~(\ref{eq:rhoBsecond}) hold for any two interacting barotropic
fluids. We now specialise to the cosmologically relevant case in which
fluid $A$ is dark energy and fluid $B$ is pressureless cold dark
matter. The specialisation does two things. It reduces the general
four-term structure of equation~(\ref{eq:rhoAsecond}) to a form whose
individual contributions can be physically interpreted in terms of
$\Lambda$CDM departures, and it removes the $\omega_B'$ term from
the dark-matter equation, isolating the dark-energy cross-coupling
as the sole dynamical source for $\rho_m''$ beyond free streaming.

\subsection{Second-order equations for dark energy and dark matter}

We set $\omega_A = \omega_{DE}$ and $\omega_B = 0$. The condition
$\omega_B = 0$ defines pressureless cold dark matter and implies
$\omega_B' = 0$ identically — not as an additional assumption but
as a consequence of the definition. Fluid $C$ remains a separately
conserved radiation component and does not appear in the equations
below. (Radiation is included in the general setup for completeness
and because it enters the $\Omega_i$-variable system discussed as
future work in Section~\ref{sec:future}; it plays no role in the
second-order equations derived here.)

Substituting into equation~(\ref{eq:rhoAsecond}):
\begin{equation}
\rho_{DE}'' = \left[9(1+\omega_{DE})^2 + 6\alpha(1+\omega_{DE})
              + \alpha^2 - 3\omega_{DE}'\right]\rho_{DE}.
\label{eq:rhoDEsecond}
\end{equation}
Substituting into equation~(\ref{eq:rhoBsecond}) with $\omega_B=0$:
\begin{equation}
\rho_m'' = 9\rho_m
           - \alpha\bigl[3(2+\omega_{DE})+\alpha\bigr]\rho_{DE}.
\label{eq:rhomatterSecond}
\end{equation}

Two features of these equations are worth noting immediately.
In equation~(\ref{eq:rhomatterSecond}), the $\omega_B'$ term has
vanished because dark matter is pressureless, leaving the self-dilution
term $9\rho_m$ and a cross-term sourced by dark energy. The coefficient
$3(2+\omega_{DE}) = 3(1+\omega_{DE})+3$ decomposes as the sum of the
dark-energy pressure contribution $3(1+\omega_{DE})$ and the
dark-matter dilution factor $3$. Consequently, even in the
cosmological-constant limit $\omega_{DE}=-1$ the coefficient does not
vanish: $3(2-1)+\alpha = 3+\alpha$, so the interaction always sources
curvature in the dark-matter trajectory, regardless of the equation
of state. Even when $\omega_{DE}'=0$, the interaction forces the
dark-matter trajectory to deviate from the pressureless free-streaming
form $\rho_m''=9\rho_m$, with a deviation growing with $|\alpha|$
and with $\rho_{DE}$.

In equation~(\ref{eq:rhoDEsecond}), the diagnostic term $-3\omega_{DE}'$
appears, as anticipated from Property~1 of Section~\ref{sec:structure}.
Both equations carry dimensions of energy density, consistent with $\rho''$
being a second derivative with respect to the dimensionless e-fold
number $N$.

\subsection{Cosmological-constant limit and the curvature diagnostic}

A cosmological constant has $\omega_{DE}=-1$ and $\omega_{DE}'=0$.
Substituting into equation~(\ref{eq:rhoDEsecond}):
\begin{equation}
\rho_{DE}'' = \alpha^2\,\rho_{DE}.
\label{eq:LambdaLimit}
\end{equation}
This is a clean and physically transparent result. When $\alpha=0$
all four terms in equation~(\ref{eq:rhoDEsecond}) vanish simultaneously,
giving $\rho_{DE}''=0$: the non-interacting cosmological constant has
a perfectly linear trajectory in e-fold time, evolving at a constant
rate. When $\alpha\neq 0$ the interaction generates non-zero curvature
$\alpha^2\rho_{DE}$ even though the equation of state is fixed. This
curvature is a purely interaction-driven effect with no analogue in
non-interacting dark-energy models. It implies that a measurement of
the curvature of the dark-energy density trajectory cannot be interpreted
as evidence for dynamical dark energy unless the interaction parameter
$\alpha$ is independently constrained.

\subsection{Hierarchical expansion and the diagnostic term}

Defining $\delta\omega \equiv 1+\omega_{DE}$ as the departure from
the cosmological-constant value, equation~(\ref{eq:rhoDEsecond})
becomes
\begin{equation}
\rho_{DE}'' = \left[\alpha^2
              + 6\alpha\,\delta\omega
              + 9\,\delta\omega^2
              - 3\omega_{DE}'\right]\rho_{DE}.
\label{eq:rhoDEexpanded}
\end{equation}
The four terms are ordered by powers of $\delta\omega$ and $\alpha$,
but the hierarchy between them depends on where in parameter space
one is operating. We discuss each term in turn and then characterise
the overall structure.

The baseline term $\alpha^2$ is the cosmological-constant-with-interaction
residual from equation~(\ref{eq:LambdaLimit}). It is present whenever
$\alpha\neq 0$, regardless of whether the equation of state departs
from $-1$.

The term $6\alpha\,\delta\omega$ is linear in both $\alpha$ and
$\delta\omega$. It is a genuine cross-term: it vanishes when either
$\alpha=0$ or $\delta\omega=0$, and is therefore present only when
both the interaction and a departure from $\omega_{DE}=-1$ are
simultaneously non-zero. Its sign depends on the sign of $\alpha\,\delta\omega$:
for quintessence-like dark energy ($\delta\omega>0$) with energy flowing
to dark matter ($\alpha>0$), this term adds to the curvature.

The term $9\,\delta\omega^2$ is second order in $\delta\omega$ and
independent of $\alpha$. It is the non-interacting curvature
contribution from a static dark-energy fluid with
$\omega_{DE}\neq -1$, and survives in the $\alpha\to 0$ limit as
the sole departure from $\rho_{DE}''=0$.

The term $-3\omega_{DE}'$ is the diagnostic term. It is first order
in the rate of change of the equation of state, independent of
$\alpha$, and has no counterpart at first order. Its sign encodes
the direction of evolution: $\omega_{DE}'<0$ (freezing, in the
Caldwell--Linder classification \cite{Caldwell2005}) enhances the curvature
relative to the static case, while $\omega_{DE}'>0$ (thawing)
suppresses it.

\begin{figure}[H]
  \centering
  \includegraphics[width=\textwidth]{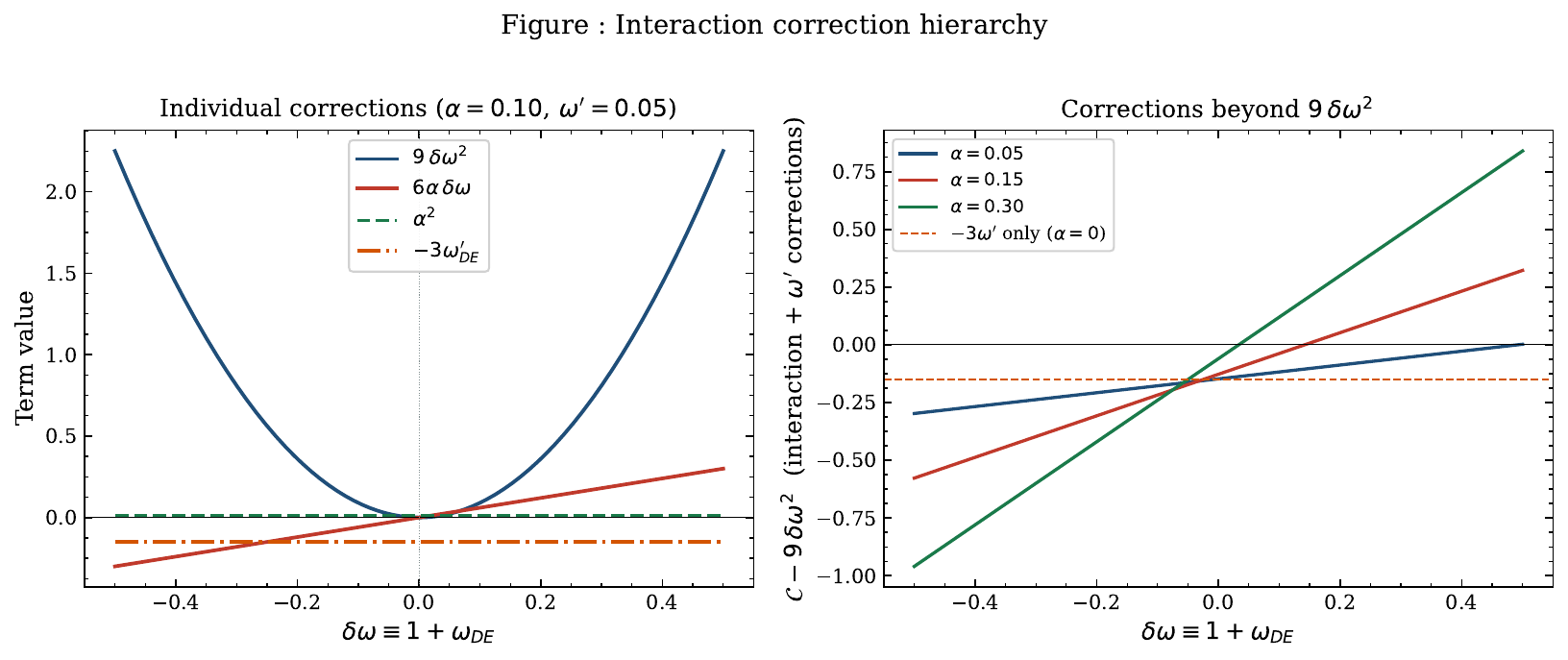}
  \caption{Interaction correction hierarchy for equation~(\ref{eq:rhoDEexpanded}).
    \textit{Left}: the four terms plotted individually as functions of
    $\delta\omega \equiv 1+\omega_{DE}$, for $\alpha=0.10$ and
    $\omega_{DE}'=0.05$. The term $9\,\delta\omega^2$ (blue) is even
    and dominates away from $\delta\omega=0$; the cross-term
    $6\alpha\,\delta\omega$ (red) is odd and changes sign at the
    $\Lambda$CDM point; the quadratic interaction $\alpha^2$ (green)
    is a positive constant; and $-3\omega_{DE}'$ (orange) is a negative
    constant that shifts the total curvature for $\omega_{DE}'>0$.
    \textit{Right}: total correction beyond $9\,\delta\omega^2$ for
    three values of $\alpha$, with the $\alpha=0$ baseline shown dashed.
    Near $\delta\omega=0$ the diagnostic term $-3\omega_{DE}'$ dominates
    regardless of $\alpha$, making it the primary signal in models
    consistent with current observational constraints on $\omega_{DE}$.}
  \label{fig:correction_hierarchy}
\end{figure}

The hierarchy among these four terms is not universal. Away from the
cosmological-constant point ($|\delta\omega|\gg 0$), the even term
$9\,\delta\omega^2$ grows quadratically and dominates; near
$\delta\omega\approx 0$, however, that term is suppressed and the
$-3\omega_{DE}'$ term becomes the leading correction beyond the
interaction baseline $\alpha^2$ for any value of $\alpha$.
This inversion is visible in Figure~\ref{fig:correction_hierarchy}
and is observationally significant: for dark-energy models consistent
with the DESI preference for $|\omega_{DE}+1|\lesssim 0.1$ \cite{DESI},
the diagnostic term $-3\omega_{DE}'$ is the dominant signal, not
the quadratic static departure $9\,\delta\omega^2$. Standard
parametric analyses that assume the CPL form
$\omega_{DE}(a) = w_0 + w_a(1-a)$ \cite{Chevallier2001,Linder2003} capture $\omega_{DE}'$
only implicitly: since $d/dN = a\,d/da$, the CPL derivative is
$\omega_{DE}' = -w_a e^N$ (derived explicitly in
Section~\ref{sec:example}), a quantity that depends on epoch
and cannot be independently constrained at a single instant. The present framework treats $w'_{\rm DE}$ as a quantity that can be inferred locally from the curvature of the dark-energy density trajectory, without requiring a global parametrisation of the equation of state.

\subsection{\label{sec:example}Worked example: CPL dark energy with interaction}
To demonstrate the diagnostic concretely, we apply equation~(\ref{eq:rhoDEexpanded}) to a CPL dark-energy model \cite{Chevallier2001,Linder2003} with parameters $w_0$ and $w_a$, and then specialise to $w_0=-0.9$, $w_a=-0.4$. These values lie within the DESI confidence region \cite{DESI} but are chosen for their proximity to the cosmological-constant point, which makes the hierarchy among the four terms in equation~(\ref{eq:rhoDEexpanded}) most transparent; they are not the DESI best-fit central values. We compare two values of the coupling, $\alpha=0.10$ and $\alpha=0.01$.

\subsubsection*{Analytic structure of the curvature coefficient}

The CPL equation of state in e-fold time is
\begin{equation}
\omega_{DE}(N) = w_0 + w_a(1-e^N),
\label{eq:CPL_omega}
\end{equation}
giving the departure from $\Lambda$CDM and its derivative as
\begin{equation}
\delta\omega(N) = (1+w_0) + w_a(1-e^N), \qquad
\omega_{DE}'(N) = -w_a\,e^N.
\label{eq:CPL_dw}
\end{equation}
The derivative $\omega_{DE}'(N)=-w_a e^N$ is independent of $\alpha$
and grows (in magnitude) with $N$, reflecting the accelerating evolution
of the CPL equation of state toward the present epoch. This independence
holds because the CPL form specifies $\omega_{DE}$ as an explicit
function of $N$ only; there is no feedback from the energy-density
evolution into $\omega_{DE}$, so $\omega_{DE}'$ is determined entirely
by $w_a$ and $N$.

Substituting equations~(\ref{eq:CPL_omega})--(\ref{eq:CPL_dw})
directly into equation~(\ref{eq:rhoDEexpanded}), the curvature
coefficient takes the analytic form
\begin{equation}
\mathcal{C}(N) = \alpha^2
  + 6\alpha\bigl[(1+w_0) + w_a(1-e^N)\bigr]
  + 9\bigl[(1+w_0) + w_a(1-e^N)\bigr]^2
  + 3w_a\,e^N.
\label{eq:CPL_C}
\end{equation}
This expression is exact and closed-form. Its structure is transparent:
the term $3w_a\,e^N = -3\omega_{DE}'$ is the diagnostic term, which
grows in magnitude as $e^N$ and dominates at late times for $|w_a|$ of
order unity; the terms in $\delta\omega$ capture static departures from
$\Lambda$CDM and depend on $N$ only through $w_a(1-e^N)$; and $\alpha^2$
is the constant interaction baseline.

To connect $\mathcal{C}(N)$ to the actual second derivative of the
dark-energy density, we integrate the first-order
equation~(\ref{eq:rhoAfirst}) for dark energy. With
$\omega_{DE}(N)=w_0+w_a(1-e^N)$ and constant $\alpha$, the integrating
factor gives
\begin{equation}
\rho_{DE}(N) = \rho_{DE,0}\,\exp\!\bigl(
  -\bigl[3(1+w_0+w_a)+\alpha\bigr]N
  + 3w_a(e^N - 1)\bigr),
\label{eq:CPL_rho}
\end{equation}
where $\rho_{DE,0}=\rho_{DE}(0)$ is the present-day dark-energy density.
Equation~(\ref{eq:CPL_rho}) is the explicit solution of the first-order
continuity equation for the CPL model; it reduces to the standard
$\rho_{DE}\propto a^{-3(1+w_0+w_a)}$ scaling at early times
($e^N\ll 1$) and to $\rho_{DE}\propto a^{-3(1+w_0)-\alpha}$ at late
times ($e^N\to 1$).

The second derivative follows directly from the definition
$\mathcal{C}=\rho_{DE}''/\rho_{DE}$:
\begin{equation}
\rho_{DE}''(N) = \mathcal{C}(N)\,\rho_{DE}(N),
\label{eq:CPL_rhopp}
\end{equation}
with $\mathcal{C}(N)$ given by equation~(\ref{eq:CPL_C}) and
$\rho_{DE}(N)$ by equation~(\ref{eq:CPL_rho}). This is the
central relation: the second derivative of the dark-energy density
at any epoch is the product of the curvature coefficient and the
density itself. Because $\rho_{DE}(N)>0$ at all epochs, the sign
of $\rho_{DE}''$ is determined entirely by $\mathcal{C}(N)$:
a negative $\mathcal{C}$ means the density trajectory is concave
(decelerating decrease), while a positive $\mathcal{C}$ means it
is convex (accelerating decrease). For the DESI-motivated CPL
parameters, $\mathcal{C}(N)<0$ throughout the trajectory
(as seen in Figure~\ref{fig:worked_example}), so
$\rho_{DE}''(N)<0$: the dark-energy density is decreasing at a
decelerating rate, consistent with a thawing equation of state.

One can verify equation~(\ref{eq:CPL_rhopp}) independently by
differentiating equation~(\ref{eq:CPL_rho}) twice with respect to
$N$. The first derivative is
\begin{equation}
\rho_{DE}'(N) = -\bigl[3(1+\omega_{DE}(N))+\alpha\bigr]\,\rho_{DE}(N),
\label{eq:CPL_rhop}
\end{equation}
which is precisely the first-order continuity
equation~(\ref{eq:rhoAfirst}) evaluated for the CPL model,
confirming consistency. Differentiating once more and using
equation~(\ref{eq:CPL_rhop}), one recovers
equation~(\ref{eq:CPL_rhopp}) with $\mathcal{C}(N)$ exactly as
in equation~(\ref{eq:CPL_C}), closing the circle.

Evaluating at $N=0$ ($a=1$, $z=0$) where $\delta\omega=1+w_0$ and
$\omega_{DE}'=-w_a$, equation~(\ref{eq:CPL_C}) reduces to
\begin{equation}
\mathcal{C}(0) = \alpha^2 + 6\alpha(1+w_0) + 9(1+w_0)^2 + 3w_a,
\label{eq:CPL_C0}
\end{equation}
which is a simple quadratic in $(1+w_0)$ shifted by the interaction
baseline $\alpha^2$ and the diagnostic contribution $3w_a$.
The inversion formula at $N=0$ is then
\begin{equation}
\omega_{DE}'\big|_{N=0} = -w_a =
\frac{\alpha^2 + 6\alpha(1+w_0) + 9(1+w_0)^2 - \mathcal{C}(0)}{3},
\label{eq:CPL_inversion}
\end{equation}
showing explicitly that once $\mathcal{C}(0)$, $\omega_{DE}(0)=w_0$,
and $\alpha$ are known, the equation-of-state derivative at today's
epoch is determined uniquely.

\subsubsection*{Numerical evaluation}

At $N=0$ (today, $z=0$) both models share $\omega_{DE}=-0.9$,
$\delta\omega=0.1$, and $\omega_{DE}'=0.40$. Substituting into equation~(\ref{eq:CPL_C0}) yields 
\begin{eqnarray}
\mathcal{C}\big|_{\alpha=0.10} = -1.04, \quad\mathcal{C}\big|_{\alpha=0.01} = -1.10.
\label{eq:example_today}
\end{eqnarray}
In both cases the diagnostic term $-3\omega_{DE}'=-1.20$ accounts
for over 100\% of the magnitude of $\mathcal{C}$ (the positive
static and interaction contributions partially offset the diagnostic
term, so $|\mathcal{C}|<|-3\omega_{DE}'|$); the interaction
and static contributions merely provide a positive offset of
$+0.16$ ($\alpha=0.10$) or $+0.10$ ($\alpha=0.01$). The
inversion formula $\omega_{DE}'=[\alpha^2+6\alpha\,\delta\omega
+9\,\delta\omega^2-\mathcal{C}]/3$ recovers $\omega_{DE}'=0.40$
exactly from both values of $\mathcal{C}$, regardless of $\alpha$.

The robustness of the diagnostic to the choice of $\alpha$ is the
central quantitative finding of this example. Because $-3\omega_{DE}'$
is independent of $\alpha$, reducing the coupling by a factor of ten
changes $\mathcal{C}$ by only $0.06$ at $N=0$ — a shift far smaller
than the diagnostic signal itself. The same pattern holds at earlier
epochs: at $N=-0.5$ ($z\approx 0.65$), where the model has drifted
into phantom territory ($\omega_{DE}=-1.057$,
$\omega_{DE}'=0.243$), the two couplings give
$\mathcal{C}=-0.70$ ($\alpha=0.01$) and $\mathcal{C}=-0.72$
($\alpha=0.10$), a difference of only $0.02$, while the diagnostic
term contributes $-3\omega_{DE}'=-0.73$ in both cases. Inversion
recovers $\omega_{DE}'=0.243$ exactly at both epochs and both
coupling values.

\begin{figure}[H]
  \centering
  \includegraphics[width=\textwidth]{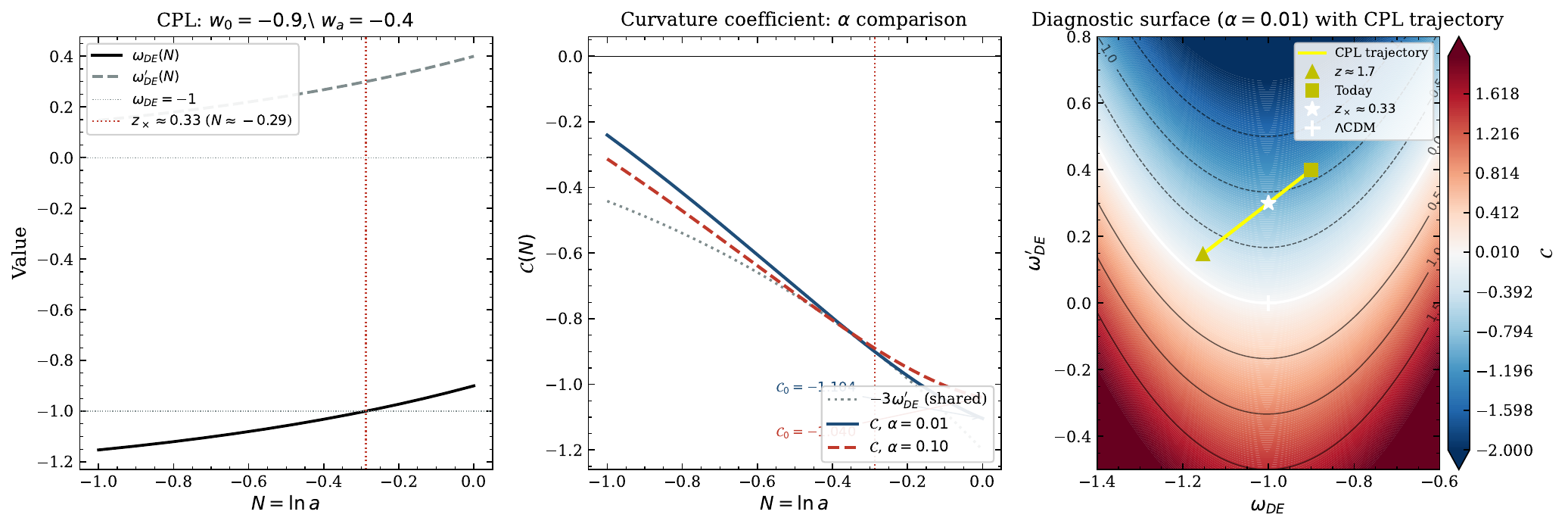}
  \caption{Worked example: CPL dark energy ($w_0=-0.9$, $w_a=-0.4$)
    for two values of the interaction coupling.
    \textit{Left}: $\omega_{DE}(N)$ and $\omega_{DE}'(N)$,
    identical for both $\alpha$ values, over $N\in[-1,0]$
    ($z\in[0,1.7]$). The model is thawing and crosses the
    phantom divide at $z_\times\approx 0.33$ (dotted vertical line),
    close to the DESI best-fit crossing near $z\approx 0.4$--$0.5$.
    \textit{Centre}: the curvature coefficient $\mathcal{C}(N)$
    for $\alpha=0.01$ (blue solid) and $\alpha=0.10$ (red dashed),
    together with the shared term $-3\omega_{DE}'$ (grey dotted).
    The two curves differ by at most $0.06$, demonstrating that
    $\mathcal{C}$ is dominated by the diagnostic term and is
    robust to the precise value of $\alpha$ within the
    phenomenologically interesting range.
    \textit{Right}: CPL trajectory on the diagnostic surface for
    $\alpha=0.01$. The triangle marks $z\approx 1.7$, the square
    marks today. At each point, $\omega_{DE}'$ is recovered
    exactly from $\mathcal{C}$ via equation~(\ref{eq:rhoDEexpanded}).}
  \label{fig:worked_example}
\end{figure}

Figure~\ref{fig:worked_example} shows the full trajectory from
$N=-1$ ($z\approx 1.7$) to $N=0$. The left panel shows
$\omega_{DE}(N)$ and $\omega_{DE}'(N)$, which are the same for
both $\alpha$ values: the model has $\omega_{DE}'>0$
throughout and crosses the phantom divide at $z_\times\approx 0.33$,
close to the $z\approx 0.4$--$0.5$ crossing preferred by DESI \cite{DESI}.
For $z>z_\times$ (i.e.\ $\omega_{DE}>-1$) the model is thawing in the
Caldwell--Linder sense; below the phantom divide the thawing/freezing
classification, which was formulated for quintessence, does not
directly apply, and we use $\omega_{DE}'>0$ as the operative
criterion for the sign of the diagnostic term.
The centre panel compares $\mathcal{C}(N)$ for $\alpha=0.01$
(blue, solid) and $\alpha=0.10$ (red, dashed), together with
the shared diagnostic term $-3\omega_{DE}'$ (grey dotted): the
two $\mathcal{C}$ curves are nearly indistinguishable, and both
track the diagnostic term closely throughout the trajectory.
The right panel shows the CPL trajectory on the diagnostic surface
for $\alpha=0.01$, confirming that the contour structure places
the trajectory firmly in the negative-$\mathcal{C}$ region and
that the inversion correctly identifies $\omega_{DE}'$ at each
point.

\subsection{\label{sec:noise}Noise propagation and detectability}

The curvature coefficient $\mathcal{C}=\rho_{DE}''/\rho_{DE}$
is a second derivative with respect to $N$, and is therefore
sensitive to the precision with which $\rho_{DE}(N)$ can be
reconstructed from $H(z)$ data. We derive a simple noise
estimate to assess whether the diagnostic signal is accessible
to near-future surveys.

Since $\rho_{DE} = 3H^2\Omega_{DE}$ and $\Omega_{DE}$ is
determined from the matter budget at each epoch, a fractional
uncertainty $\sigma_H/H$ in the Hubble rate propagates to
\begin{equation}
\frac{\sigma_{\rho_{DE}}}{\rho_{DE}} = 2\,\frac{\sigma_H}{H}
\equiv \epsilon.
\label{eq:rho_err}
\end{equation}
We note that reconstructing $\rho_{DE}$ from $H(z)$ data also requires
subtracting the matter and radiation contributions, introducing additional
uncertainties from $\Omega_m$ and $\Omega_r$ that are not propagated here.
The estimate below should therefore be regarded as a lower bound on the
full uncertainty.
Approximating the second derivative by a centred second difference
over bins of width $\Delta N$,
\begin{equation}
\rho_{DE}'' \approx
\frac{\rho_{DE}(N+\Delta N) - 2\rho_{DE}(N)
+ \rho_{DE}(N-\Delta N)}{\Delta N^2},
\label{eq:second_diff}
\end{equation}
the squared coefficients $1^2+(-2)^2+1^2=6$ give
$\sigma_{\rho_{DE}''} = \sqrt{6}\,\epsilon\,\rho_{DE}/\Delta N^2$
(assuming uncorrelated measurements). Adding the normalisation
uncertainty in quadrature, the total uncertainty on $\mathcal{C}$ is
\begin{equation}
\sigma_\mathcal{C} \approx
\frac{\sqrt{6}\,\epsilon}{\Delta N^2}
+ |\mathcal{C}|\,\epsilon.
\label{eq:sigma_C}
\end{equation}
The first term, which scales as $\epsilon/\Delta N^2$, dominates
for the bin widths relevant to spectroscopic surveys; the second
term is a minor correction of order $|\mathcal{C}|\epsilon\lesssim 0.03$
for $\epsilon\lesssim 0.02$.

For the CPL model of Section~\ref{sec:example} the signal is
$|-3\omega_{DE}'|=1.2$ at $N=0$. The signal-to-noise ratio is
\begin{equation}
\mathrm{SNR} = \frac{|-3\omega_{DE}'|}{\sigma_\mathcal{C}}.
\label{eq:SNR}
\end{equation}
At $\Delta N=0.5$ (corresponding to redshift bins $\Delta z\approx 0.5$
near $z=0$, as expected from a survey like Euclid or DESI at its
design sensitivity \cite{DESI}), and treating measurements at the three
stencil points as uncorrelated (an optimistic assumption given that
neighbouring $H(z)$ measurements from BAO or cosmic chronometers are
typically correlated), the estimates are:
\begin{eqnarray}
\sigma_H/H = 1.0\% \quad &\Rightarrow& \quad \sigma_\mathcal{C}\approx 0.22,
\quad \mathrm{SNR}\approx 5.5, \nonumber\\
\sigma_H/H = 1.5\% \quad &\Rightarrow& \quad \sigma_\mathcal{C}\approx 0.33,
\quad \mathrm{SNR}\approx 3.6.
\label{eq:SNR_numbers}
\end{eqnarray}
These estimates are order-of-magnitude upper bounds on the true SNR.
They neglect correlations between neighbouring redshift bins, covariance
introduced by the reconstruction of $H(z)$, survey window functions,
systematic uncertainties associated with numerical differentiation, and
the additional contribution from matter-subtraction uncertainties noted
above. A comprehensive assessment of detectability will require mock
survey analyses using realistic covariance matrices and reconstruction
pipelines.
Both are comfortably above the SNR$\,=3$ threshold, indicating that
the diagnostic is accessible to surveys achieving sub-2\% Hubble
rate precision over bins of this width. Current cosmic chronometer
measurements already reach $\sigma_H/H\approx 1$--$2\%$ in several
redshift bins \cite{Seikel2012}, and future spectroscopic surveys are
projected to approach $\sigma_H/H\approx 0.5\%$ \cite{DESI}.

\begin{figure}[H]
  \centering
  \includegraphics[width=\textwidth]{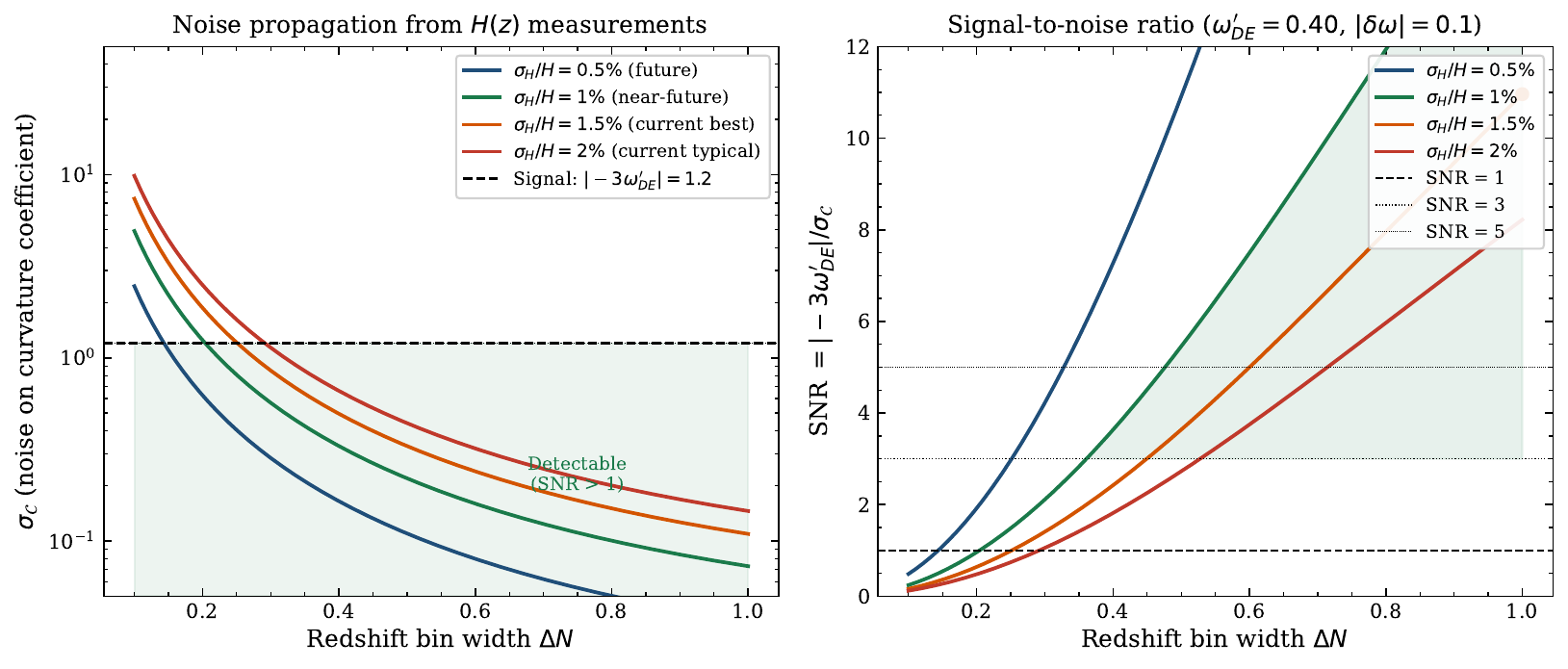}
  \caption{Noise propagation estimate for the curvature diagnostic.
    \textit{Left}: the uncertainty $\sigma_\mathcal{C}$
    (equation~\ref{eq:sigma_C}) as a function of the redshift
    bin width $\Delta N$, for four values of the fractional
    Hubble rate uncertainty $\sigma_H/H$. The dashed horizontal
    line marks the signal level $|-3\omega_{DE}'|=1.2$ for the
    worked example; the shaded region indicates
    $\sigma_\mathcal{C}<$ signal (detectable).
    \textit{Right}: signal-to-noise ratio as a function of
    $\Delta N$ for the same four precision levels. Filled circles
    mark the bin widths achieving maximum SNR within $\Delta N\leq 0.8$
    for the $\sigma_H/H=1\%$ and $1.5\%$ cases. At $\Delta N=0.5$
    the SNR exceeds 3 for $\sigma_H/H\leq 1.5\%$, placing the
    diagnostic within reach of near-future $H(z)$ surveys.}
  \label{fig:noise}
\end{figure}

The dominant term in equation~(\ref{eq:sigma_C}) scales as
$1/\Delta N^2$: wider bins reduce the noise rapidly, at the cost
of temporal resolution. A practical strategy is to use wide bins
($\Delta N\sim 0.5$) for a global detection of the diagnostic
signal, then narrow bins to resolve the epoch-dependence of
$\omega_{DE}'(N)$. At $\Delta N=0.5$ and $\sigma_H/H=0.5\%$
(the projected precision of future spectroscopic surveys \cite{DESI}),
equation~(\ref{eq:sigma_C}) gives $\sigma_\mathcal{C}\approx 0.11$
and $\mathrm{SNR}\approx 11$. Figure~\ref{fig:noise} illustrates
$\sigma_\mathcal{C}$ and the SNR as functions of $\Delta N$ for
four values of $\sigma_H/H$. We emphasise that
equation~(\ref{eq:sigma_C}) is an order-of-magnitude upper-bound
estimate; a full detectability analysis, accounting for correlated
measurements, survey window functions, and systematic errors in
the $H(z)$ reconstruction, is deferred to future work.

\subsection{\label{sec:alpha_degen}Degeneracy between $\alpha$
and $\omega_{DE}'$}

Equation~(\ref{eq:rhoDEexpanded}) shows that $\mathcal{C}$
depends on both $\alpha$ and $\omega_{DE}'$. The inversion
formula $\omega_{DE}' = [\alpha^2 + 6\alpha\,\delta\omega
+ 9\,\delta\omega^2 - \mathcal{C}]/3$ recovers
$\omega_{DE}'$ exactly when $\alpha$, $\omega_{DE}$, and
$\mathcal{C}$ are all known. In practice, $\alpha$ is itself
uncertain and must be constrained by additional observations.
We discuss how this degeneracy can be broken.

The three unknowns $\{\alpha,\,\omega_{DE},\,\omega_{DE}'\}$
at a given epoch can in principle be constrained by measuring
$\mathcal{C}$ at multiple epochs, since $\omega_{DE}'(N)$ is
a smooth function that determines the trajectory on the
diagnostic surface of Figure~\ref{fig:diagnostic_surface}.
A single epoch gives one equation~(\ref{eq:rhoDEexpanded})
in three unknowns; two epochs with independent $\omega_{DE}$
measurements give two equations and can constrain two of the
three, with $\alpha$ determined if the functional form of
$\omega_{DE}(N)$ is assumed. A model-independent approach
requires at least three epochs.

However, as the worked example of Section~\ref{sec:example}
demonstrates, the practical impact of $\alpha$ uncertainty
is small when $\alpha$ is small and $|\delta\omega|\lesssim 0.1$.
The terms involving $\alpha$ in equation~(\ref{eq:rhoDEexpanded})
contribute $\alpha^2 + 6\alpha\,\delta\omega\lesssim 0.01 + 0.06
= 0.07$ for $\alpha\leq 0.10$ and $|\delta\omega|\leq 0.1$,
compared to the diagnostic signal $|-3\omega_{DE}'|\sim 1.2$.
An uncertainty of $\pm 0.05$ in $\alpha$ propagates to an
uncertainty of $|\partial\mathcal{C}/\partial\alpha|\,\sigma_\alpha
= |2\alpha + 6\,\delta\omega|\,\sigma_\alpha
\lesssim (0.20+0.06)\times 0.05 = 0.013$ in $\mathcal{C}$,
which is negligible compared to the measurement noise
$\sigma_\mathcal{C}\sim 0.2$--$0.3$ estimated above.
The diagnostic is therefore robust to $\alpha$ uncertainty
at the level of precision achievable with near-future surveys,
provided $\alpha$ is constrained to within $\pm 0.05$ by
independent probes such as growth-rate or perturbation-level
measurements. Existing analyses of large-scale structure and CMB
data already constrain linear couplings of this form to
$|\alpha|\lesssim 0.05$--$0.10$ at $1\sigma$ \cite{Wang2016,Giare2024},
so the required precision on $\alpha$ is consistent with current
observational bounds.

\section{\label{sec:discussion}Discussion}

The derivations and examples of Sections~\ref{sec:derivation}
and~\ref{sec:de} establish that the curvature of the dark-energy
density trajectory in e-fold time is a complementary dynamical diagnostic :
it carries information about $\omega_{DE}'$ that the slope of the
same trajectory does not, and it does so independently of any assumed
parametric form for the equation of state. This section discusses
how the result fits into the existing theoretical landscape
(Section~\ref{sec:wplane}), what the current observational evidence
implies (Section~\ref{sec:obscontext}), and which extensions are
most tractable (Section~\ref{sec:future}).

\subsection{\label{sec:wplane}Relation to the $w$-$w'$ plane}

Caldwell and Linder \cite{Caldwell2005} introduced the plane spanned by
$(\omega_{DE},\,\omega_{DE}')$ to classify quintessence models without
reference to a specific scalar-field potential. In their framework,
models with $\omega_{DE}'>0$ are \textit{thawing} — the equation of
state was formerly frozen near $-1$ and is now rolling away from it —
while models with $\omega_{DE}'<0$ are \textit{freezing} — the
equation of state is decelerating toward $-1$; for quintessence
this means approaching from above ($\omega_{DE}>-1$), while
phantom models with $\omega_{DE}<-1$ have $\omega_{DE}'>0$
and are thawing away from the phantom divide. This
classification has been widely applied to holographic dark energy,
phantom, quintom, and interacting scenarios \cite{Bahamonde2018}.

The present paper grounds that classification in the continuity
equations rather than in a specific potential, and extends it to the
interacting setting. In the non-interacting limit $\alpha\to 0$,
equation~(\ref{eq:rhoDEexpanded}) reduces to
\begin{equation}
\rho_{DE}'' = \bigl[9\,\delta\omega^2 - 3\omega_{DE}'\bigr]\rho_{DE}
\qquad (\alpha = 0),
\label{eq:noninteracting_limit}
\end{equation}
which is precisely the curvature form of the Caldwell--Linder
classification: the even term $9\,\delta\omega^2$ is the static
departure from $\Lambda$CDM, and the term $-3\omega_{DE}'$ encodes
the thawing/freezing character. Equation~(\ref{eq:noninteracting_limit})
serves as a consistency check on the general result and makes explicit
which structure is inherited from the non-interacting case and which
is new.

Once $\alpha\neq 0$, two additional terms appear: the cross-term
$6\alpha\,\delta\omega$, which vanishes at the $\Lambda$CDM point, and
the baseline $\alpha^2$, which does not. The $\alpha^2$ term means that
an interacting cosmological constant produces non-zero curvature in
the density trajectory — a feature entirely outside the Caldwell--Linder
framework. The diagnostic surface in
Figure~\ref{fig:diagnostic_surface} is the interacting analogue of the
$w$-$w'$ plane, with $\mathcal{C}=\rho_{DE}''/\rho_{DE}$ as the
observable axis; comparing the $\alpha=0$ and $\alpha=0.15$ panels
shows directly how the interaction baseline $\alpha^2$ shifts all
contours upward while leaving the overall structure intact.

The present diagnostic is complementary to other higher-order cosmological diagnostics proposed in the literature, such as the Statefinder hierarchy, jerk-based diagnostics, and the $Om$ diagnostic\cite{Sahni2003,Arabsalmani2011,Sahni2008}. Those diagnostics are constructed from successive derivatives of the expansion factor or the Hubble parameter and are primarily designed to distinguish between competing cosmological models at the background level. By contrast, the curvature coefficient
\[
\mathcal{C}=\frac{\rho_{\rm DE}''}{\rho_{\rm DE}}
\]
arises directly from differentiating the dark-energy continuity equation and therefore probes the local evolution of the dark-energy density itself. Its explicit dependence on $w'_{\rm DE}$ follows from energy conservation rather than from an assumed scalar-field potential or phenomenological parametrisation. The diagnostic should therefore be viewed as complementary to existing geometric diagnostics, providing an alternative perspective on the dynamics of the dark sector rather than replacing established approaches.
\subsection{\label{sec:obscontext}Observational context}

The observational case for $\omega_{DE}'\neq 0$ has strengthened
considerably since the DESI DR1 and DR2 data releases \cite{DESI},
which combined with CMB and supernova data show a preference for
evolving dark energy over $\Lambda$CDM. Analysing the DR1 results,
Cort\`es and Liddle \cite{Cortes2024} showed that the evidence is
concentrated in $w'$ rather than in the mean offset of $w$ from $-1$:
the pivot equation of state is consistent with $-1$, but $w'$ is
non-zero at moderate significance. This is precisely the quantity
$\omega_{DE}'$ that enters equation~(\ref{eq:rhoDEsecond}) exclusively
at second order, making the DESI results direct observational motivation
for the second-order approach developed here.

Within interacting models the picture is further complicated because
the effective equation of state inferred from background observables
differs from the microscopic $\omega_{DE}$ \cite{Avelino2009,Yang2018}: energy
transfer modifies the background evolution so that a microscopically
static dark energy can mimic a dynamical one at the level of distance
measurements, and vice versa. The term $-3\omega_{DE}'$ in
equation~(\ref{eq:rhoDEsecond}) responds to the microscopic rate of
change, not the effective value, and therefore provides a handle on
this degeneracy that purely geometric probes cannot break.

The observable quantity is $\mathcal{C}=\rho_{DE}''/\rho_{DE}$,
which can in principle be reconstructed from second differences of the
dark-energy density history $\rho_{DE}(z)$, itself recoverable from
$H(z)$ data via the Friedmann equation. Non-parametric Gaussian process
reconstructions of $H(z)$ and the dark-energy density have been
demonstrated on current data \cite{Seikel2012,Escamilla2023}, and the interaction
kernel has similarly been reconstructed without parametric assumptions
\cite{Escamilla2023}. As shown in Figure~\ref{fig:diagnostic_surface}, a
measurement of $\mathcal{C}=\rho_{DE}''/\rho_{DE}$ at known $\alpha$
constrains the system to a horizontal line in the
$(\omega_{DE},\omega_{DE}')$ plane; combined with an independent
measurement of $\omega_{DE}$, it recovers $\omega_{DE}'$ uniquely
through the inversion $\omega_{DE}' = [\alpha^2 + 6\alpha\,\delta\omega
+ 9\,\delta\omega^2 - \mathcal{C}]/3$. For the range
$|\omega_{DE}'|\sim 0.05$--$0.10$ suggested by current data
\cite{Cortes2024}, the diagnostic signal $|3\omega_{DE}'|\sim 0.15$--$0.30$
is comparable to or larger than the static departure $9\,\delta\omega^2\sim 0.09$
for $|\delta\omega|=0.1$, placing it within reach of near-future
$H(z)$ reconstructions. A full noise analysis connecting second
differences of $H(z)$ to uncertainties in $\mathcal{C}$ is deferred
to future work.

\subsection{\label{sec:future}Directions for future work}

Three directions follow directly from the present analysis. The most
tractable is a full critical-point analysis of the system rewritten
in the expansion-normalised variables $\Omega_i$ of
Section~\ref{sec:setup}. As established in
Section~\ref{sec:structure}, moving to $\Omega_i$ variables brings
in the Raychaudhuri equation and globally couples all three components.
In those variables the term $-3\omega_{DE}'$ modifies the Jacobian
at the fixed points of the first-order system, potentially shifting
their stabilities and changing which late-time attractors are
physically accessible. This analysis would make precise contact with
the thawing and freezing classification of \cite{Caldwell2005} in the interacting
setting, extending the Caldwell--Linder programme to include $\alpha$.

\begin{figure}[H]
  \centering
  \includegraphics[width=\textwidth]{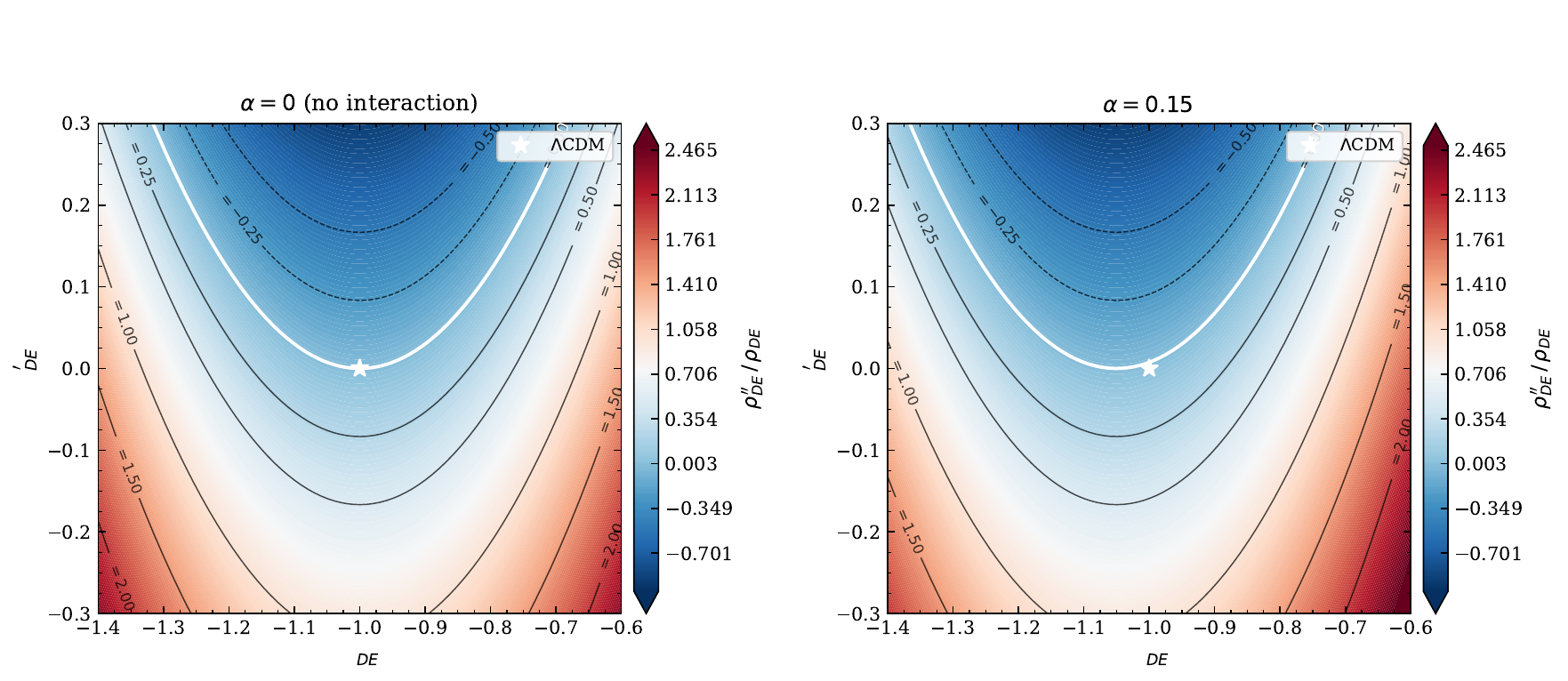}
  \caption{Diagnostic surface: the curvature coefficient
    $\mathcal{C}(\omega_{DE},\,\omega_{DE}')$ as a two-dimensional
    colour map, for $\alpha=0$ (\textit{left}) and $\alpha=0.15$
    (\textit{right}). Black contours give labelled values of
    $\mathcal{C}$; the white contour marks $\mathcal{C}=0$ (the
    trajectory inflection locus); the white star marks the $\Lambda$CDM
    point $(\omega_{DE},\omega_{DE}')=(-1,0)$. At fixed $\alpha$, a
    measurement of $\mathcal{C}$ constrains the system to a horizontal
    line in this plane; combined with an independent measurement of
    $\omega_{DE}$, it determines $\omega_{DE}'$ uniquely via
    $\omega_{DE}' = [\alpha^2 + 6\alpha\,\delta\omega + 9\,\delta\omega^2
    - \mathcal{C}]/3$. The interaction shifts all contours upward by
    $\alpha^2$ relative to the $\alpha=0$ panel but preserves the
    diagnostic structure.}
  \label{fig:diagnostic_surface}
\end{figure}

A second direction is to extend the interaction
ansatz~(\ref{eq:Qansatz}) to non-linear forms $Q_{AB}(\rho_A,\rho_B,H)$
and to explore models with $\alpha<0$, corresponding to energy transfer
from dark matter to dark energy. Changing the sign of $\alpha$ reverses
the cross-term in equation~(\ref{eq:rhomatterSecond}) and shifts the
contours in Figure~\ref{fig:diagnostic_surface}, but
the presence of $-3\omega_{DE}'$ and its independence from $\alpha$
are unaffected: the structural argument of
Section~\ref{sec:derivation} holds for any smooth interaction, and
$\omega_{DE}'$ will appear at second order in all cases.

The third direction is to connect the present framework directly to
the DESI evidence \cite{DESI,Cortes2024,Giare2024} by constraining
$\omega_{DE}'$ from reconstructed density histories without assuming
a CPL or other parametric form. The diagnostic surface in
Figure~\ref{fig:diagnostic_surface} maps out the structure of this
constraint: a measurement of $\mathcal{C}$ at known $\alpha$ confines
the system to a horizontal line in the $(\omega_{DE},\omega_{DE}')$
plane, and a joint measurement of $\omega_{DE}$ then determines
$\omega_{DE}'$ uniquely. Taken together, the three directions chart
a path from the analytic results of the present paper toward a
model-independent, non-parametric test of dynamical dark energy in
the interacting dark-sector setting.

Although the present analysis has been developed for linear interactions of the form
\[
Q=\alpha\rho H,
\]
within a spatially flat FLRW spacetime containing barotropic fluids, the central structural result is considerably more general. The explicit appearance of $w'_{\rm DE}$ follows from differentiating the dark-energy continuity equation and therefore reflects the temporal evolution of the equation of state rather than the specific choice of interaction model. Alternative interaction prescriptions will modify the algebraic form of the curvature coefficient and may introduce additional terms, but the emergence of derivatives of the equation of state at second order is expected to be a generic consequence of differentiating the continuity equations. Establishing the precise form of these corrections for more general interaction kernels remains an interesting direction for future investigation.
\section{\label{sec:conclusions}Conclusions}

We have derived the second-order evolution equations for the energy
densities of two interacting barotropic fluids in a flat FLRW
spacetime, with a linear energy-transfer coupling $Q_{AB}=\alpha\rho_A H$.
The main results are as follows.

\textit{(i) Structural result.} The time derivative $\omega_{DE}'$ of the
dark-energy equation of state enters the second-order continuity equation
through the term $-3\omega_{DE}'\rho_{DE}$. This term is structurally
absent from the first-order equations, where $\omega_{DE}$ appears only
as a coefficient of the density. Two dark-energy models that agree on
$\omega_{DE}$ at a given instant but differ in $\omega_{DE}'$ are locally
indistinguishable at first order and distinguishable at second order. The second-order equations therefore make explicit information about the evolution of the equation of state that is only implicitly encoded in the first-order formulation.

\textit{(ii) Curvature diagnostic.} The curvature coefficient
$\mathcal{C}=\rho_{DE}''/\rho_{DE}$ defines a model-independent
diagnostic for dynamical dark energy. In the cosmological-constant limit
($\omega_{DE}=-1$, $\omega_{DE}'=0$), $\mathcal{C}=\alpha^2$: any
non-zero curvature in this limit is interaction-driven, not dynamical.
Departures from $\omega_{DE}=-1$ generate corrections hierarchically
ordered in $\delta\omega\equiv 1+\omega_{DE}$ and $\alpha$. Near
the cosmological-constant point, which is consistent with current DESI
constraints \cite{DESI,Cortes2024}, the diagnostic term
$-3\omega_{DE}'$ dominates over the static departure $9\,\delta\omega^2$
for any value of $\alpha$.

\textit{(iii) Non-interacting limit.} Setting $\alpha\to 0$ recovers
the Caldwell--Linder curvature structure $[9\,\delta\omega^2
- 3\omega_{DE}']\rho_{DE}$ \cite{Caldwell2005}, providing a consistency check
and grounding the thawing/freezing classification in the continuity
equations. The $\alpha^2$ term represents a genuinely new feature
absent from the non-interacting framework: an interacting cosmological
constant has non-zero density curvature.

\textit{(iv) Worked example and noise estimate.} Applied to a
DESI-motivated CPL model ($w_0=-0.9$, $w_a=-0.4$), the diagnostic
recovers $\omega_{DE}'$ exactly at all epochs for both $\alpha=0.01$
and $\alpha=0.10$. A simple noise propagation from $H(z)$ uncertainties
gives SNR$\,\approx 5.5$ at $\sigma_H/H=1\%$ and $\Delta N=0.5$,
rising to SNR$\,\approx 11$ at $\sigma_H/H=0.5\%$. The diagnostic
signal dominates over the uncertainty in $\alpha$ for $\alpha\lesssim 0.10$
at survey precisions $\sigma_H/H\leq 1.5\%$.

\textit{(v) Future directions.} Three extensions follow directly:
a critical-point analysis of the second-order system in
expansion-normalised variables, an extension to non-linear interaction
ansätze, and a full non-parametric constraint on $\omega_{DE}'$ from
reconstructed $H(z)$ histories applied to DESI data
\cite{DESI,Cortes2024,Giare2024}.

\acknowledgments

The author acknowledges funding from the University of Cape Town's NGP programme.



\begin{thebibliography}{99}
\bibitem{Frieman2008}J.~A.~Frieman, M.~S.~Turner, and D.~Huterer,``Dark Energy and the Accelerating Universe,''\emph{Annu. Rev. Astron. Astrophys.} \textbf{46}, 385--432 (2008),doi:10.1146 annurev.astro.46.060407.145243.
\bibitem{Weinberg1989} S.~Weinberg,``The Cosmological Constant Problem,''Rev. Mod. Phys. \textbf{61}, 1--23 (1989).
\bibitem{Peebles2003}P.~J.~E.~Peebles and B.~Ratra, ``The Cosmological Constant and Dark Energy,''Rev. Mod. Phys. \textbf{75}, 559--606 (2003).
\bibitem{Copeland2006} E.~J.~Copeland, M.~Sami, and S.~Tsujikawa,``Dynamics of Dark Energy,'' Int. J. Mod. Phys. D \textbf{15}, 1753--1936 (2006).
\bibitem{Weinberg2013} D.~H.~Weinberg \emph{et al.}, ``Observational Probes of Cosmic Acceleration,''Phys. Rep. \textbf{530}, 87--255 (2013).
\bibitem{DESI}DESI Collaboration (Adame~A~G et al) 2025 DESI 2024 VI: cosmological constraints from the measurements of baryon acoustic oscillations, \textit{J.\ Cosmol.\ Astropart.\ Phys.} \textbf{2025}(02) 021
\bibitem{Amendola2000}L.~Amendola,``Coupled Quintessence,''Phys.\ Rev.\ D \textbf{62}, 043511 (2000).
\bibitem{Zimdahl2001}W.~Zimdahl, D.~Pav\'on, and L.~P.~Chimento,``Interacting Quintessence,''Phys.\ Lett.\ B \textbf{521}, 133--138 (2001).
\bibitem{Wang2016}B.~Wang, E.~Abdalla, F.~Atrio-Barandela, and D.~Pav\'on,``Dark Matter and Dark Energy Interactions: Theoretical Challenges, Cosmological Implications and Observational Signatures,''Rep.\ Prog.\ Phys.\ \textbf{79}, 096901 (2016).
\bibitem{Bolotin2015}Y.~L.~Bolotin, A.~Kostenko, O.~A.~Lemets, and D.~A.~Yerokhin,``Cosmological Evolution With Interaction Between Dark Energy and Dark Matter,''Int.\ J.\ Mod.\ Phys.\ D \textbf{24}, 1530007 (2015).
\bibitem{Valiviita2008}J.~Valiviita, E.~Majerotto, and R.~Maartens,``Instability in Interacting Dark Energy and Dark Matter Fluids,''JCAP \textbf{07}, 020 (2008).
\bibitem{Kunz2009}M.~Kunz,``The Dark Degeneracy: On the Number and Nature of Dark Components,''Phys.\ Rev.\ D \textbf{80}, 123001 (2009).
\bibitem{Copeland1998}E.~J.~Copeland, A.~R.~Liddle, and D.~Wands,``Exponential Potentials and Cosmological Scaling Solutions,''Phys.\ Rev.\ D \textbf{57}, 4686--4690 (1998).
\bibitem{Wainwright1997}J.~Wainwright and G.~F.~R.~Ellis,\emph{Dynamical Systems in Cosmology}(Cambridge University Press, Cambridge, 1997).
\bibitem{Coley2003}A.~A.~Coley,\emph{Dynamical Systems and Cosmology}(Kluwer Academic Publishers, Dordrecht, 2003).
\bibitem{Bahamonde2018}S.~Bahamonde,et al ``Dynamical Systems Applied to Cosmology: Dark Energy and Modified Gravity,''Phys.\ Rep.\ \textbf{775--777}, 1--122 (2018).
\bibitem{Caldwell2005}R.~R.~Caldwell and E.~V.~Linder,``The Limits of Quintessence,''Phys.\ Rev.\ Lett.\ \textbf{95}, 141301 (2005).
\bibitem{Linder2006}E.~V.~Linder,``The Dynamics of Quintessence, The Quintessence of Dynamics,''Gen.\ Relativ.\ Gravit.\ \textbf{40}, 329--356 (2008) [arXiv:0704.2064 [astro-ph]].
\bibitem{Scherrer2008}R.~J.~Scherrer and A.~A.~Sen,``Thawing Quintessence with a Nearly Flat Potential,''Phys.\ Rev.\ D \textbf{77}, 083515 (2008).
\bibitem{Weinberg1972}S.~Weinberg,\emph{Gravitation and Cosmology: Principles and Applications of the General Theory of Relativity}(Wiley, New York, 1972).
\bibitem{Mukhanov2005}V.~Mukhanov,\emph{Physical Foundations of Cosmology}(Cambridge University Press, Cambridge, 2005).
\bibitem{Sahni2003}V.~Sahni, T.~D.~Saini, A.~A.~Starobinsky and U.~Alam,``Statefinder---a new geometrical diagnostic of dark energy,''\emph{JETP Lett.} \textbf{77}, 201--206 (2003),doi:10.1134/1.1574831.
\bibitem{Arabsalmani2011} M.~Arabsalmani and V.~Sahni,``The Statefinder hierarchy: An extended null diagnostic for concordance cosmology,''\emph{Phys. Rev. D} \textbf{83}, 043501 (2011),doi:10.1103/PhysRevD.83.043501.
\bibitem{Sahni2008}V.~Sahni, A.~Shafieloo and A.~A.~Starobinsky,``Two new diagnostics of dark energy,''\emph{Phys. Rev. D} \textbf{78}, 103502 (2008),doi:10.1103/PhysRevD.78.103502.

\bibitem{Yang2018}W.~Yang et al,``Interacting Dark Energy with Time-Varying Equation of State and the $H_0$ Tension,''
Phys.\ Rev.\ D \textbf{98}, 123527 (2018).
\bibitem{Giare2024}
W.~Giare, M.~A.~Sabogal, R.~C.~Nunes and E.~Di~Valentino,
``Interacting Dark Energy after DESI Baryon Acoustic Oscillation Measurements,''
Phys.\ Rev.\ Lett.\ \textbf{133}, 251003 (2024).

\bibitem{Avelino2009}
P.~P.~Avelino, L.~M.~G.~Beca, C.~J.~A.~P.~Martins and P.~Pinto,
``Is $w\neq-1$ Evidence for a Dynamical Dark Energy Equation of State?''
Phys.\ Rev.\ D \textbf{80}, 067302 (2009).

\bibitem{OsanoOreta2019}
B.~Osano and T.~Oreta,``Multi-Fluid Theory and Cosmology: A Convective Variational Approach to Interacting Dark Sectors,'' Int.\ J.\ Mod.\ Phys.\ D \textbf{28}, 1950078 (2019).

\bibitem{Osano2025}
B.~Osano,``Dark Energy: A Dynamical Systems Approach to the Reconstruction of the Equation of State,''J.\ Math.\ Phys.\ \textbf{66}, 092501 (2025).

\bibitem{Osano2025b}B.~Osano,``Cosmological Evolution: A Study of Transition Periods,'' J.\ Numer.\ Simul.\ Phys.\ Math.\ \textbf{1}, 67--75 (2025).

\bibitem{Cortes2024}
M.~Cort\'es and A.~R.~Liddle,
``Interpreting DESI's Evidence for Evolving Dark Energy,'' JCAP \textbf{12}, 007 (2024).

\bibitem{Chevallier2001}
M.~Chevallier and D.~Polarski,
``Accelerating Universes with Scaling Dark Matter,''
Int.\ J.\ Mod.\ Phys.\ D \textbf{10}, 213--224 (2001).

\bibitem{Linder2003}
E.~V.~Linder,
``Exploring the Expansion History of the Universe,''
Phys.\ Rev.\ Lett.\ \textbf{90}, 091301 (2003).

\bibitem{Seikel2012}
M.~Seikel, C.~Clarkson and M.~Smith,
``Reconstruction of Dark Energy and Expansion Dynamics Using Gaussian Processes,''
JCAP \textbf{06}, 036 (2012).

\bibitem{Escamilla2023}
L.~A.~Escamilla, O.~Akarsu, E.~Di~Valentino and J.~A.~V\'azquez,
``Model-Independent Reconstruction of the Interacting Dark Energy Kernel:
Binned and Gaussian Process Approaches,''
JCAP \textbf{11}, 051 (2023).
\end{thebibliography}
\end{document}